\documentstyle[sprocl,epsfig]{article}

\def\Journal#1#2#3#4{{#1} {\bf #2}, #3 (#4)}

\def\PLB{{\em Phys. Lett.}  B}
\def\PRL{\em Phys. Rev. Lett.}

\def\ZPC{{\em Z. Phys.} C}

\def\be{\begin{equation}}
\def\ee{\end{equation}}
\def\bea{\begin{eqnarray}}
\def\eea{\end{eqnarray}}
\def\bp{{\bf p}}
\def\bq{{\bf q}}
\def\ds{\displaystyle}
\newcommand{\bcen}{\begin{center}}
\newcommand{\ecen}{\end{center}}

\begin{document}

\title{SEPARATION BETWEEN SOURCES OF PIONS AND PROTONS IN CENTRAL 
AU+AU COLLISIONS AT THE AGS (E877).}
\author{D. MI\'SKOWIEC}
\address{Gesellschaft f\"ur Schwerionenforschung mbH \\
Planckstr.~1, 
D-64291 Darmstadt, Germany\\
E-mail: D.Miskowiec@gsi.de}

\maketitle\abstracts{
Two-particle pion-pion, proton-proton, and pion-proton correlations 
were measured in central Au+Au collisions at 11~GeV/c per nucleon. 
The data were analyzed using one- and three-dimensional correlation 
functions. 
The pion-proton correlation functions exhibit asymmetry which indicates 
that the sources of protons and pions are separated in the direction 
along the accelerator beam. }

\section{Introduction}
The two-particle correlation function, defined as the ratio between 
the two-particle density and the product of single particle densities 
\be
\label{c1}
C(\bp_1,\bp_2)  =
\frac{\ds
\sigma \frac{d^6\sigma}{d^3p_1 d^3p_2}
}{\ds
\frac{d^3\sigma}{d^3p_1} \frac{d^3\sigma}{d^3p_2}} \ ,
\ee
contains important information about the space-time-momentum 
distribution at the time of last interaction (freeze-out) 
of particles produced in a nuclear collision.~\cite{boa90} 
The correlations of identical pions, 
dominated by the Bose-Einstein statistics and by mutual Coulomb interaction, 
exhibit a peak at $\bq:=\bp_2-\bp_1=0$, the width of which 
is inversely proportional to the source size. 
Analysis of $C$ as a function of different components of the momentum 
difference yields the source dimensions in different directions. 
The two-proton correlation is a product of the Fermi statistics as well as 
the mutual Coulomb and strong interactions. This correlation shows a 
peak at the relative momentum $q\approx~40~$MeV/c, 
the amplitude of which is roughly inversely proportional 
to the volume of the proton source. 
The unlike-pions and pion-proton correlations are generated mostly by the 
Coulomb interaction and thus have an exponential-like shape at $\bq=0$. 
Similarly as for identical pions, even if with somewhat lower sensitivity, 
the width of this peak depends on the average r.m.s. distance 
between the particles at freeze-out. 

\section{Experiment and data analysis}
I present here results of the correlation analysis of particles 
produced in central Au+Au collisions at 11~GeV/c per nucleon. 
Five millions central (4\% of the geometrical cross section) 
collision events were recorded in Fall 1994 by the E877 Collaboration at the 
AGS.  
The experimental setup, presented in Figure~1, consisted of 
three calorimeters, used to determine the centrality and 
the orientation of the reaction plane and to generate the centrality trigger, 
and 
a forward spectrometer for measuring momenta of pions and protons. 
The momentum resolution of the spectrometer was $\Delta p/p\approx 3\%$. 
The acceptance allowed simultaneous measurement of positive 
and negative particles (Figure~2). 
\begin{figure}[b!]
\scalebox{0.6}{\includegraphics{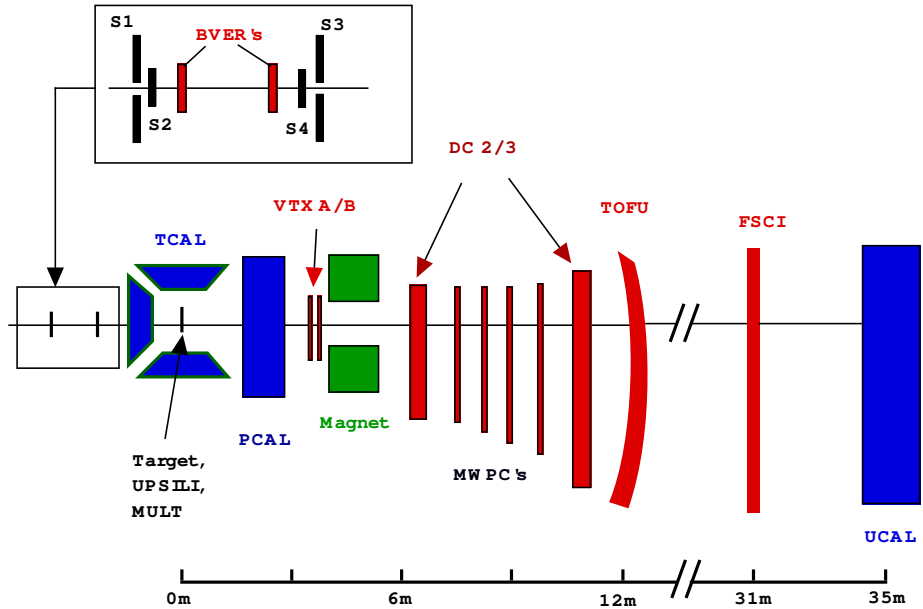}}

\vspace{-3cm}
\caption{
The E877 setup at the AGS. Three calorimeters, TCAL, PCAL, and 
UCAL, were used to determine the collision centrality and the orientation of 
the reaction plane. The particles used in the correlation analysis were 
measured in the forward spectrometer. }
\end{figure}

\begin{figure}[t!]
\vspace*{-1cm}
\hspace*{-1cm}
\scalebox{0.8}{\includegraphics{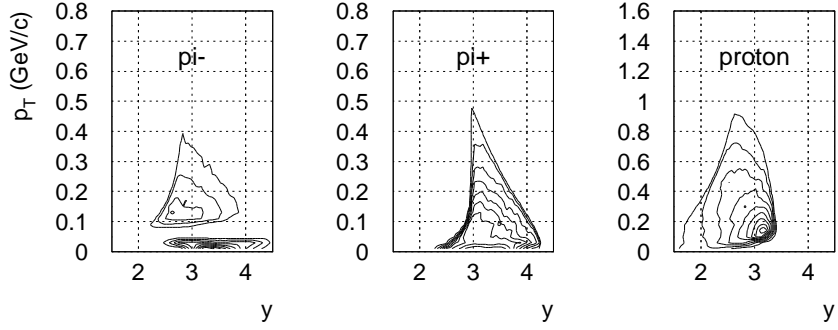}}
\vspace{-1cm}
\caption{
Distribution of analyzed particles in transverse momentum and rapidity. 
The target and the beam rapidities were 0 and 3.1, respectively. 
The hole in the acceptance for negative pions was caused by the accelerator 
beam going through the detectors. }
\end{figure}

\begin{figure}[b!]
\vspace*{-1cm}
\scalebox{0.8}[0.55]{\includegraphics{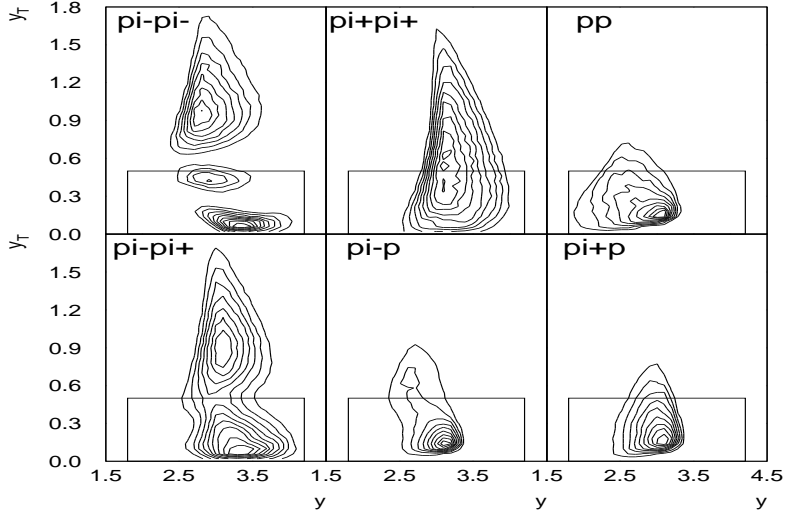}}
\caption{
Distribution of accepted pairs. 
Pair rapidity $y$ is defined as $\frac{1}{2}\log((E+P_z)/(E-P_z))$, 
with $E$ and $P_z$ being the pair energy and longitudinal momentum 
in the LAB frame. 
The pair transverse rapidity $y_{\rm T}$ is defined as 
$\frac{1}{2}\log((E^*+P_{\rm T})/(E^*-P_{\rm T}))$ 
where $E^*=\sqrt{E^2-P_z^2}$ is 
the pair energy in the longitudinally comoving system (LCMS), and 
$P_{\rm T}$ is the pair transverse momentum. 
The rectangle represents the analysis cut. }
\end{figure}
\clearpage

The numerators of the correlation functions were generated using pairs of 
identified particles detected in the forward spectrometer. 
The distributions of the analyzed particles and particle pairs are 
shown in Figures~2 and 3, respectively. 
The denominators were obtained by event mixing. 
We mixed only events with similar centralities and similar orientations of 
the reaction plane. 
The two track resolution problems were avoided by rejecting from the numerator 
and the denominator all track pairs which did not have enough separation in 
the tracking detectors. 
The correlation functions were normalized to unity at the flat part 
between $q=100~$MeV/c and $q=400~$MeV/c. 
No Coulomb correction was applied. 
The relative momentum $\bq$ was evaluated in the pair c.m.s. 
which is the natural choice when analyzing correlation functions of 
particles with different masses. 
The $q_{\rm x}, q_{\rm y}$, and $q_{\rm z}$ components are defined as follows: 
$\hat{\bf z}$ is the accelerator beam axis, 
$\hat{\bf x}$ is in the reaction plane, 
%($\hat{\bf x}\parallel{\bf b}$, where ${\bf b}$ is the impact parameter 
%pointing from the center of the target to the center of the projectile 
%nucleus), 
and $\hat{\bf y}$ is perpendicular to the beam and to the reaction plane. 
The reaction plane is determined from the azimuthal distribution of the 
transverse energy. 
The sign of $x$ is defined such that the transverse flow of forward protons 
is in the positive-$x$ direction. 

\section{Results}
The obtained one-dimensional correlation functions are presented 
in Figure~4. 
The correlation functions of identical pions are dominated 
by the positive Bose-Einstein correlation, slightly suppressed by the 
repulsive Coulomb interaction. The bumb in the two-proton correlation is 
caused by the attractive strong interaction, and the minimum at low 
relative momenta is caused by the Coulomb interaction and the Fermi 
statistics. The non-identical particle correlations are mainly due to 
Coulomb interaction. The little peak in the $\pi^-$--proton correlation 
function is caused by $\Lambda^0$ decays. The position and the width 
of this peak was used as a check of the momentum calibration and of the 
knowledge of the momentum resolution. 

The line represents the result of a calculation described in the next section.
The calculation overestimates both pion-proton correlation functions while 
all other correlations are reasonably reproduced. 

\begin{figure}[ht!]
\hspace*{-1cm}
\scalebox{0.8}{\includegraphics{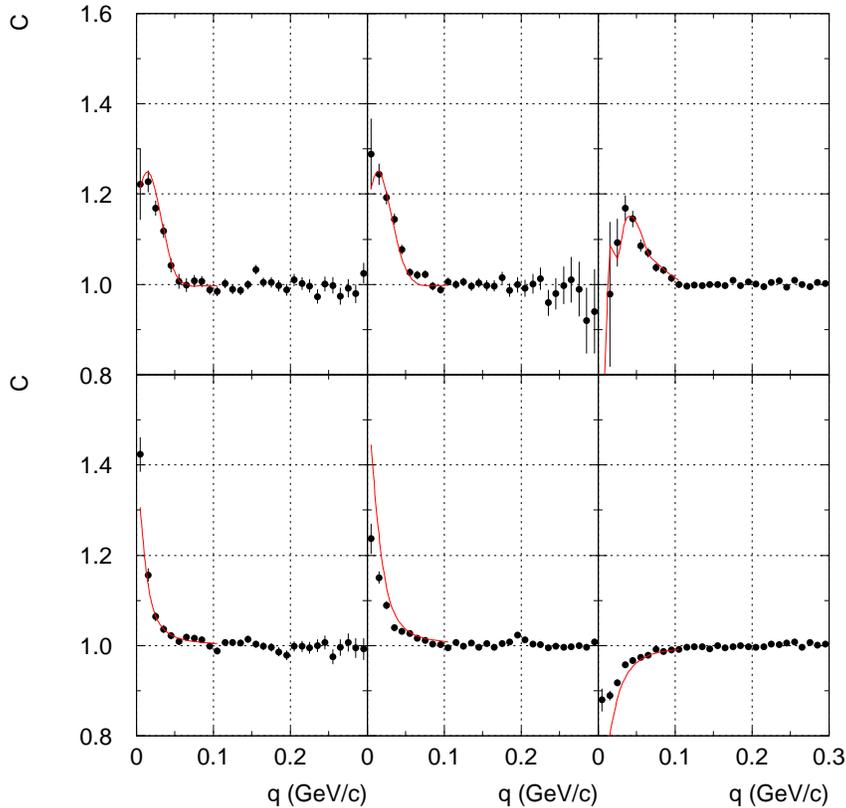}}

\vspace{-0.5cm}
\caption{
One-dimensional correlation functions vs. relative momentum 
in the pair c.m.s. }
\end{figure}

The projections of the three-dimensional correlation functions are shown 
in Figure~5. The projections were done while restricting the other two 
components of $\bq$ to the range from -50~MeV/c to 50~MeV/c. 
The line, again, represents the calculations. 
Inspection of Figure~5 allows to trace back 
the origin of the disagreement between the data and the calculation 
in the pion-proton correlations to the asymmetric shape 
of the peak in the data, not reproduced by the calculation. 

%%%%%%%%%%%%%%%%%%%%%%%%% 3-dim correlations %%%%%%%%%%%%%%%%%%%%%%%%%%%%%%%%%
\begin{figure}[ht!]
\vspace*{-1cm}
\scalebox{0.6}{\includegraphics{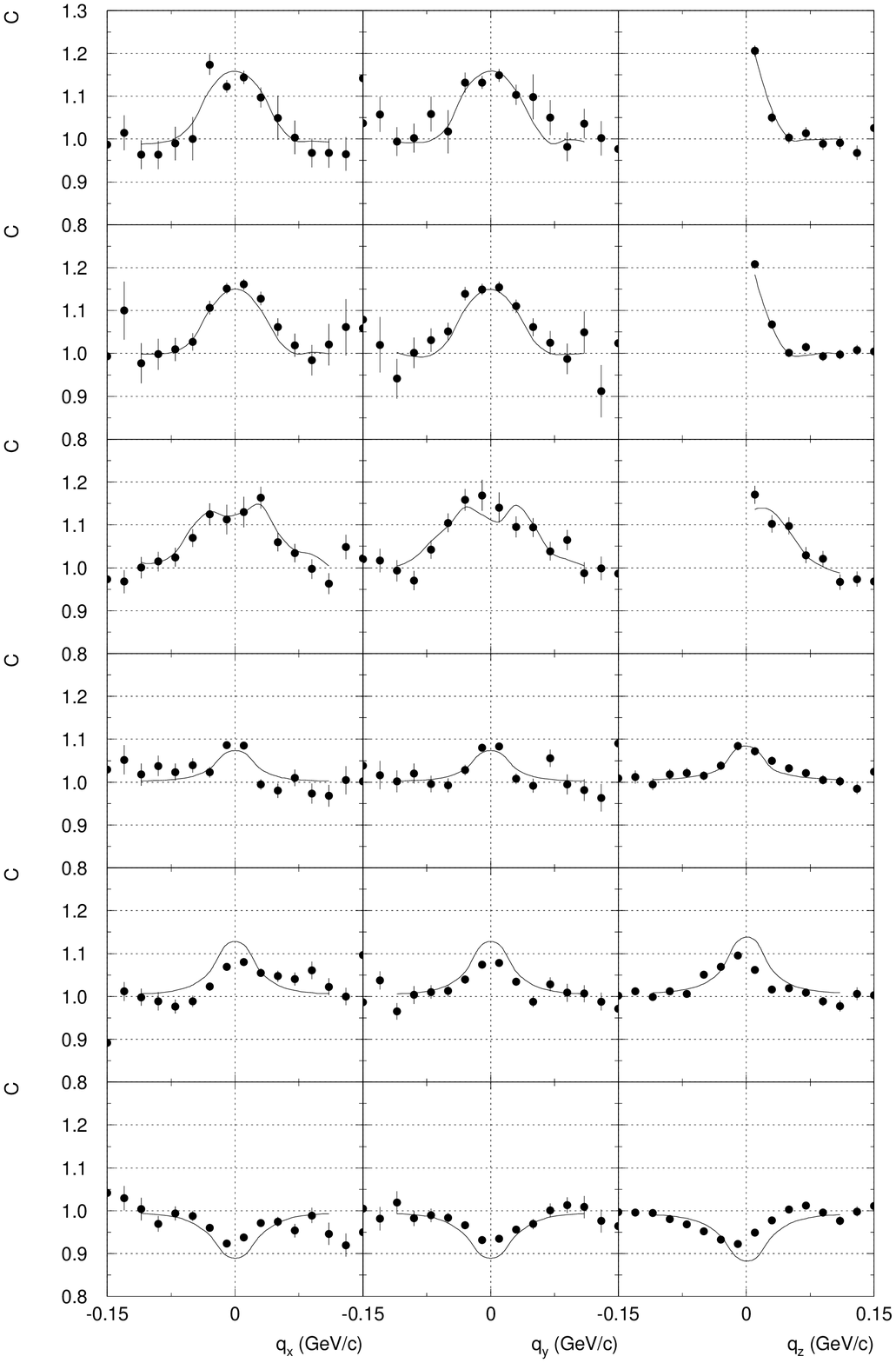}}

\vspace{-0.5cm}
\caption{
Projections of three-dimensional correlation functions.}
\vspace{-16cm}
\hspace*{11cm}
\begin{minipage}[t]{4cm}
\it \normalsize

$\pi^- \ \pi^-$\\

\vspace{1.75cm}
$\pi^+ \ \pi^+$\\

\vspace{1.75cm}
$p \ p$\\

\vspace{1.75cm}
$\pi^+ \ \pi^-$\\

\vspace{1.75cm}
$\pi^- \ p$\\

\vspace{1.75cm}
$\pi^+ \ p$\\

\vspace{1cm}
\end{minipage}
\end{figure}
\vfill
%%%%%%%%%%%%%%%%%%%%%%%%% 3-dim correlations -  end %%%%%%%%%%%%%%%%%%%%%%%%%%%

As observed by Lednicky {\it et al}~\cite{led96}, an asymmetric peak 
in the correlation function between non-identical particles is a sign 
of spatial or temporal separation between the sources of the two 
particle species. 
(The mechanism of this effect was explained by B.~Erazmus in her talk 
during this conference.) 
Since the asymmetry in our data is in the $q_z$ direction of 
both pion-proton correlations, we conclude that there is a finite 
separation between the sources of pions and protons in the beam direction. 
The sign of the asymmetry indicates that at freeze-out protons 
are more forward than pions with similar (forward) rapidities. 
In order to estimate the size of this displacement we performed the 
calculation described below.

\section{Calculation}
We generated pairs of particles. Each particle was described by four 
position and four momentum coordinates. 
The position coordinates were generated according to a double-gaussian 
distribution for pions (core-halo model~\cite{cso96}) 
and a single-gaussian for protons:
\bea
g^{\rm pion}({\bf r}) & = &
f \ \exp\left(-\frac{x^2+y^2}{2 R_{\rm T}^2}-
               \frac{z^2}{2 R_{\rm z}^2}\right) \ + (1-f) \ 
\exp\left(-\frac{r^2}{2 R_{\rm halo}^2}\right) \\
g^{\rm proton}({\bf r}) & = &
\exp\left(-\frac{x^2+y^2}{2 R_{\rm T}^2}-
               \frac{z^2}{2 R_{\rm z}^2}\right)
\eea
The parameter $f$ represents the fraction of pions coming from the core, 
i.e. not from long-lived resonances. 
It is related to the correlation strength $\lambda$ by 
\be
f^2=\lambda \ .
\ee

The momenta were sampled from the experimental momentum distributions. 
The Lednicky code~\cite{led82,led94} was used to calculate a weight for 
every pair. The code accounts for the (anti)symmetrization effect as 
well as for the Coulomb and strong interactions. 
Subsequently, the momenta were smeared by the experimental momentum 
resolution and the two-track separation cut was applied. 
Finally, every pair was processed through the same analysis as the 
experimental data, except that the Lednicky's weight and unity were used to 
increment the histograms for the numerator and the denumerator of the 
correlation function, respectively (no event mixing). 

The parameters of the gaussian position distributions were adjusted 
to fit the data. 
The three-dimensional proton-proton correlation function allows to 
determine the transverse and longitudinal proton source size, 
even if the coupling between the direction of \bq{} and ${\bf r}$ is not 
as clean as it is for pions. 
The best agreement was obtained for 
$R^{\rm proton}_{\rm T}=4.75$~fm and 
$R^{\rm proton}_{\rm z}=3.25$~fm. 
The $\pi^-\pi^-$, $\pi^+\pi^+$, and $\pi^+\pi^-$ correlations could be 
simultaneously reproduced with $f=0.72$, 
$R^{\rm pion}_{\rm T}=5.0$~fm, 
$R^{\rm pion}_{\rm z}=7.8$~fm, and
$R^{\rm pion}_{\rm halo}=30$~fm. 
The correlation functions calculated using these parameters are represented 
by lines in Figures 4 and 5. 
The pion-proton correlations, calculated using the same parameter values 
and assuming that the pion and proton sources are centered at the same 
point, disagree with the data. 

\begin{figure}[h!]
\vspace{-0.5cm}
\hspace*{-1cm}
\scalebox{0.65}{\includegraphics{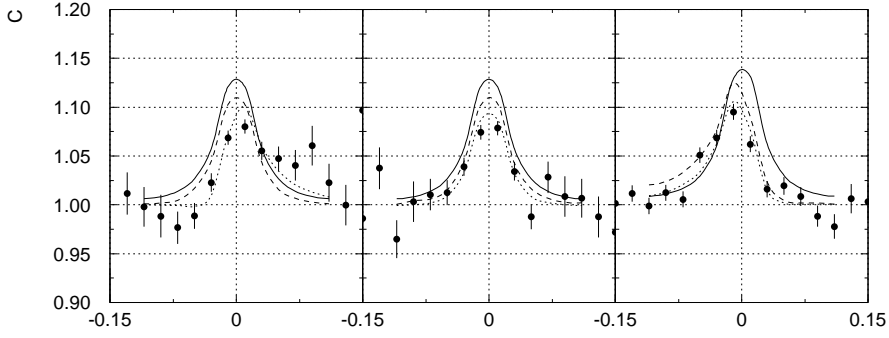}}
\vspace{-1.2cm}
\caption{
Projections of three-dimensional $\pi^-p$ correlation functions.
The solid, dashed, and dotted lines represent calculations with 
($\Delta x=0$~fm, $\Delta z=0$~fm), 
($\Delta x=0$~fm, $\Delta z=10$~fm), and 
($\Delta x=-10$~fm, $\Delta z=10$~fm), respectively. 
$\Delta {\bf r}$ is defined 
as the difference between the positions of the sources of protons and pions:
$\Delta {\bf r} = 
\langle {\bf r}^{\rm proton} \rangle - \langle {\bf r}^{\rm pion} \rangle$. 
}
\end{figure}
\begin{figure}[ht!]
\vspace{-0.5cm}
\hspace*{-1cm}
\scalebox{0.65}{\includegraphics{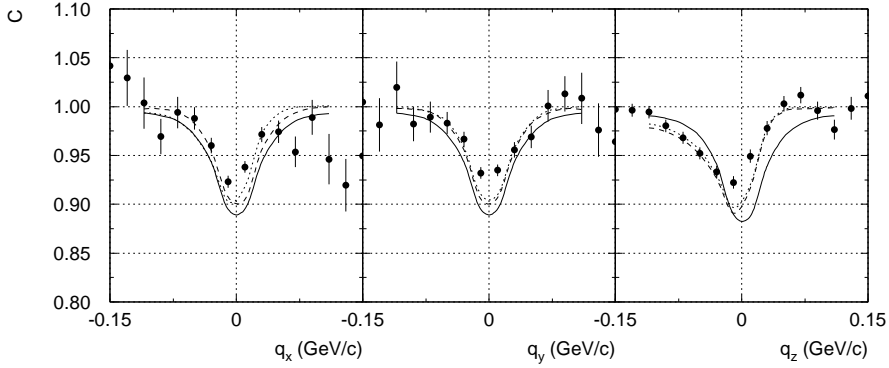}}
\vspace{-1cm}
\caption{
Projections of three-dimensional $\pi^+p$ correlation functions.
The solid, dashed, and dotted lines represent calculations with 
($\Delta x=0$~fm, $\Delta z=0$~fm), 
($\Delta x=0$~fm, $\Delta z=10$~fm), and 
($\Delta x=5$~fm, $\Delta z=10$~fm), respectively. 
}
\end{figure}
The discrepancy can be removed by assuming a finite displacement 
between the sources of pions and protons.  
Calculations using different values of the displacement are compared 
to the data in Figures~6 and 7. 
A reasonable agreement was achieved with a 10~fm separation in $z$ 
and -10~fm or +5~fm, depending on the charge of the pions, in $x$. 
\clearpage

\section{Conclusion}
The corresponding source distributions are shown in Figure 8. 
The proton source is located 10~fm more forward than the sources of pions. 
The $\pi^-$ source is, in addition, shifted in the reaction plane 
by 10~fm in the direction of positive $x$, i.e. in the direction of 
the flow of the forward protons. 
The $\pi^+$ source is shifted by 5~fm towards negative $x$. 
\begin{figure}[ht!]
\vspace{-4cm}
\hspace*{-1cm}
\scalebox{0.7}{\includegraphics{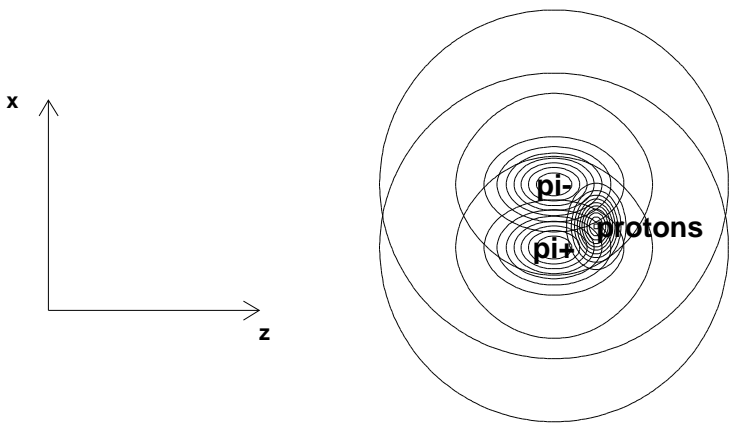}}
\vspace{-5cm}
\caption{
The $\pi^-$, $\pi^+$, and proton source distributions yielding the 
best fit to the experimental correlation functions. }
\end{figure}

A similar displacement of the pion and proton sources was observed 
in the RQMD generated events.~\cite{vol97} 
In hydrodynamical models 
the displacement in $z$ is a natural consequence of longitudinal expansion 
combined with the mass difference between pions and protons.~\cite{hei,cso} 
However, the magnitude of the observed displacement seems large compared to 
the size of the system and the expected duration of the emission. 
Indeed, even assuming that the pion source stays at the origin (in the 
nucleus-nucleus c.m. system), and the proton source moves with beam 
rapidity, it would take $\tau_f=10$~fm/c proper time to reach the separation 
of 10~fm. During this time the system would expand to the total length of 
\be
L=2 \sinh (\frac{y_{\rm B}}{2}) \ \tau_f \approx 50~{\rm fm} \ .
\ee
Since it hard to imagine that the pion source does not move at all, 
this value can be considered as the lower limit. 

In summary, asymmetry in non-identical particle correlation functions 
was observed for the first time. 
The asymmetry indicates a displacement between the sources of pions and 
protons. 
The magnitude of this displacement is related to the total duration of 
the reaction.


\section*{References}
\begin{thebibliography}{99}
\bibitem{boa90} 
D.~H.~Boal, C-K.~Gelbke, and B.~K.~Jennings, {\em Rev. Mod. Phys.}{62}{553}
{1990}.
\bibitem{led96} 
R.~Lednicky {\it et al}, \Journal{\PLB}{373}{30}{1996}. 
\bibitem{baska} 
B.~Erazmus, this conference. 
\bibitem{cso96}
T.~Cs\"org\H{o}, B.~L\"orstad, and J.~Zim\'anyi, \Journal{\ZPC}{71}{491}{1996}.
\bibitem{led82}
R.Lednicky, V.L.Lyuboshitz, Sov.J.Nucl.Phys. 35 (1982) 770. 
\bibitem{led94}
R.Lednicky {\it et al}, SUBATECH-94-22, Nantes).
\bibitem{vol97}
S.~Voloshin, R.~Lednicky, S.~Panitkin, and Nu~Xu, \Journal{\PRL}{79}{4766}{1997}. 
\bibitem{hei}
U.~Heinz, private communication. 
\bibitem{cso}
T.~~Cs\"org\H{o}, private communication. 
\end{thebibliography}
\end{document}